\documentclass[12pt]{iopart}

\usepackage{epsfig}
\usepackage{rotating}

\begin{document}

\title[Final results of experiment to search for 2$\beta$ processes in zinc and tungsten ...]
{Final results of experiment to search for 2$\beta$ processes in zinc and tungsten
with the help of radiopure ZnWO$_4$ crystal scintillators}

\author{P.~Belli$^1$,
R.~Bernabei$^{1,2}$\footnote{Corresponding author. E-mail: rita.bernabei@roma2.infn.it},
F.~Cappella$^{3,4}$, R.~Cerulli$^5$, F.A.~Danevich$^6$, S.~d'Angelo$^{1,2}$,
A.~Incicchitti$^{3,4}$, V.V.~Kobychev$^6$, D.V.~Poda$^{5,6}$, V.I.~Tretyak$^6$}

\address{$^1$ INFN sezione Roma ``Tor Vergata'', I-00133 Rome, Italy}
\address{$^2$ Dipartimento di Fisica, Universit\`a di Roma ``Tor Vergata'', I-00133, Rome, Italy}
\address{$^3$ INFN sezione Roma ``La Sapienza'', I-00185 Rome, Italy}
\address{$^4$ Dipartimento di Fisica, Universit\`a di Roma ``La Sapienza'', I-00185 Rome, Italy}
\address{$^5$ INFN, Laboratori Nazionali del Gran Sasso, I-67100 Assergi (AQ), Italy}
\address{$^6$ Institute for Nuclear Research, MSP 03680 Kyiv, Ukraine}

\begin{abstract}

A search for the double beta decay of zinc and tungsten isotopes has been performed
with the help of radiopure ZnWO$_4$ crystal
scintillators ($0.1-0.7$~kg) at the Gran Sasso National Laboratories of the INFN.
The total exposure of the low background measurements is 0.529 kg $\times$ yr.
New improved half-life limits on the double beta decay modes of $^{64}$Zn, $^{70}$Zn, $^{180}$W, and $^{186}$W
have been established at the level of $10^{18}-10^{21}$~yr. In particular,
limits on double electron capture and electron capture with positron emission in
$^{64}$Zn have been set:
$T_{1/2}^{2\nu2K}\geq1.1\times 10^{19}$~yr,
$T_{1/2}^{0\nu2\varepsilon}\geq3.2\times 10^{20}$~yr,
$T_{1/2}^{2\nu\varepsilon\beta^+}\geq9.4\times 10^{20}$~yr, and
$T_{1/2}^{0\nu\varepsilon\beta^+}\geq8.5\times 10^{20}$~yr,  all at 90\% C.L.
Resonant neutrinoless double electron capture in $^{180}$W has been restricted on
the level of $T_{1/2}^{0\nu2\varepsilon}\geq1.3\times 10^{18}$~yr.
A new half-life limit on $\alpha$ transition of $^{183}$W to the metastable
excited level 1/2$^-$ 375 keV of $^{179}$Hf has been established:
$T_{1/2}\geq6.7\times 10^{20}$~yr.

\end{abstract}


\vspace{2pc}

ZnWO$_4$ crystal scintillators. Low background experiment

\maketitle

\section{Introduction}

Double beta (2$\beta$) processes are nuclear transformations when
the charge of nuclei changes by two units: $\left(A, Z\right)
\rightarrow \left(A, Z\pm2 \right)$. There are two main modes of
2$\beta$ decay: two neutrino mode ($2\nu$) when two neutrinos are
emitted together with two beta particles, 
and neutrinoless mode ($0\nu$). $0\nu2\beta$ decay
violates the lepton number by two units and therefore is forbidden
in the Standard Model (SM) \cite{2bRev1}. However, the
$0\nu2\beta$ decay is predicted in some SM extensions where
neutrino is expected to be a true neutral particle equivalent to
its antiparticle (Majorana particle) \cite{2bRev2}. Experiments on
neutrino oscillations already gave evidence for neutrino to be
massive \cite{Oscil}, however these experiments are sensitive only
to the differences of squared masses of neutrinos. The observation
of $0\nu2\beta$ decay could resolve important problems of particle
physics: what is the absolute scale of neutrino mass? Which
neutrino mass hierarchy (normal, inverted, or quasi-degenerate) is
realized in nature? Is the neutrino Majorana ($\nu =
\overline{\nu}$) or Dirac ($\nu \neq \overline{\nu}$) particle? Is
the lepton number absolutely conserved? Additionally,
investigations of neutrinoless double $\beta$ decay could test
admixture of right-handed currents in electroweak interaction and
existence of majorons\footnote{Massless or light bosons that arise
due to a global breakdown of $\left(B - L\right)$ symmetry, where
$B$ and $L$ are the baryon and the lepton number, respectively.
Literature considers $0\nu2\beta$ decay channels with one
($0\nu2\beta M1$) \cite{Moh91,Ber92}, two ($0\nu2\beta M2$)
\cite{Moh88,Bam95}, and "bulk" ($0\nu2\beta bM$) \cite{Moh00}
majoron emissions.}.

While $2\nu2\beta$ decay is allowed in the SM, it is a second
order process in perturbation theory characterized by extremely
low probability. Investigations of the $2\nu2\beta$
decay examine theoretical calculations of the nuclear matrix
elements, contributing to the development of theoretical
description of $0\nu2\beta$ decay.

Double beta decay experiments are concentrated mainly on $2\beta$
processes with emission of two electrons ($2\beta^-$). Two
neutrino mode of $2\beta^-$ decay was detected for 11 nuclides
among 35 candidates; corresponding half-lives are in the range of
$10^{18}-10^{24}$~yr \cite{2bTable,Bar10a,Bar10b,Ack11}. In
addition, the $2\nu2\beta^-$ transitions of $^{100}$Mo and
$^{150}$Nd to the first $0^+$ excited states of daughter nuclei
were observed too \cite{2bTable,Bar10a,Bar10b}. To-date the
$2\nu2\beta^-$ decay is the rarest radioactive decay ever
discovered. Developments in the experimental techniques during
last two decades lead to impressive improvement of sensitivity to
the neutrinoless mode of 2$\beta^-$ decay up to the level of
$T_{1/2} \sim 10^{23}-10^{25}$~yr \cite{2bTable,Bar10b}. Moreover,
some possible positive indication for $^{76}$Ge with
$T^{0\nu2\beta}_{1/2} = 2.2 \times 10^{25}$~yr has been mentioned
in \cite{Kla06}, and new experiments are in preparation both on
$^{76}$Ge \cite{GERDA,Majorana} and other isotopes.

A more modest sensitivity was reached in the experiments searching
for $2\beta$ processes with decreasing charge of nuclei: capture
of two electrons from atomic shells (2$\varepsilon$), electron
capture with positron emission ($\varepsilon\beta^+$), and double
positron decay ($2\beta^+$). There are 34 possible candidates for
2$\varepsilon$ capture; among them, only 22 and 6 nuclei can also
decay through $\varepsilon\beta^+$ and $2\beta^+$ channels,
respectively \cite{2bTable}. In contrast to the $2\beta^-$ decay,
even the allowed $2\nu$ mode of 2$\varepsilon$,
$\varepsilon\beta^+$, and $2\beta^+$ processes are still not
detected in direct experiments\footnote{For completeness, we
remind that a possible evidence of $2\nu2\varepsilon$ capture in
$^{130}$Ba with $T^{2\nu2\varepsilon}_{1/2} \approx
\left(0.5-2.7\right) \times 10^{21}$~yr has been reported in
geochemical studies \cite{Mes01,Puj09}.} and the obtained
half-life limits are much more modest. The most sensitive
experiments have given limits on the 2$\varepsilon$,
$\varepsilon\beta^+$, and $2\beta^+$ processes at the level of
$10^{18}-10^{21}$~yr \cite{2bTable,Bar10b}. Reasons for such a
situation are: 1) lower energy releases ($Q_{\beta\beta}$) in
comparison with those in $2\beta^-$ decay, that results in lower
probabilities of the processes\footnote{The value of half-life is
inversely related to the phase-space factor ($G$); the latter
depends on energy release as $G \sim Q^{11}_{\beta\beta}$ for
$2\nu2\beta$ decay and $\sim Q^{5}_{\beta\beta}$ for $0\nu2\beta$
decay \cite{2bRev2}.}, as well as provides difficulties to
suppress background; 2) usually lower natural abundances
($\delta$) of $2\beta^+$ isotopes (which are typically lower than
5\% with only a few exceptions\footnote{Only 6 nuclides from a
complete list of 34 isotopes-candidates on 2$\varepsilon$,
$\varepsilon\beta^+$, and $2\beta^+$ processes have natural
abundances of more than 5\% \cite{2bTable}.}). Nevertheless,
studies of 2$\varepsilon$ and $\varepsilon\beta^+$ decays are
important, because observation of neutrinoless mode of such
process could help to distinguish between the mechanisms of
neutrinoless $2\beta$ decay (is it due to non-zero neutrino mass
or to the right-handed admixtures in weak interactions)
\cite{Hir94}.

Zinc tungstate (ZnWO$_4$) scintillators contain four potentially $2\beta$ active isotopes:
$^{64}$Zn, $^{70}$Zn, $^{180}$W, and $^{186}$W (see Table~1). It is worth to mention, $^{64}$Zn and $^{186}$W have
comparatively large natural abundance that allows to apply ZnWO$_4$ detectors without high cost
enriched isotopes. Moreover, the $2\nu2\beta^-$ decay of $^{186}$W is expected to be strongly suppressed
\cite{Cas94}, that could provide favorable conditions to search for neutrinoless
$2\beta^-$ decays, including processes with emission of majoron(s) which have broad energy spectra,
somewhat similar to that of the two neutrino mode. The $^{180}$W isotope is also an interesting $2\beta$ nuclide
because, in case of the capture of two electrons from the $K$ shell ($E_K = 65.4$~keV), the decay energy is rather small
(13 $\pm$ 4)~keV. Such a coincidence could give a resonant enhancement of the 0$\nu$ double electron
capture to the corresponding level of the daughter nucleus \cite{Win55,Vol82,Ber83,Suj02,Suj04,Suj04,Kri11}.

\begin{table}[htb]
\caption{Potentially 2$\beta$  active isotopes of zinc and tungsten present in ZnWO$_4$ crystal scintillators.}
\begin{center}
\begin{tabular}{|l|l|l|l|l|}
\hline Transition             & Energy release      & Isotopic                  & Decay             & Number of mother \\
~                   & ($Q_{\beta\beta}$), keV \cite{Aud03}    & abundance           & channels      & nuclei in 100~g of \\
~                             & ~                                   & (\%) \cite{Ber11} & ~                     & ZnWO$_4$ crystal \\
\hline
$^{64}$Zn $\to$ $^{64}$Ni     & 1095.7(0.7)             & 49.17(75)         & $2\varepsilon$, $\varepsilon\beta^+$ & $9.45\times10^{22}$ \\
\hline
$^{70}$Zn $\to$ $^{70}$Ge     & 998.5(2.2)              & 0.61(10)          & $2\beta^-$    & $1.17\times10^{21}$ \\
\hline
$^{180}$W $\to$ $^{180}$Hf    & 144(4)                  & 0.12(1)           & $2\varepsilon$ & $2.31\times10^{20}$ \\
\hline
$^{186}$W $\to$ $^{186}$Os    & 489.9(1.4)              & 28.43(19)         & $2\beta^-$   & $5.47\times10^{22}$ \\
\hline
\end{tabular}
\end{center}
\end{table}

The best to-date half-life limits on different modes and channels
of $2\beta$ processes in zinc and tungsten isotopes (except of
$0\nu2\beta^-$ decays of $^{186}$W) were obtained in previous
stages of this experiment \cite{Bel08,Bel09}. The best half-life
limits on $0\nu2\beta^-$ decays of $^{186}$W to the ground and
 excited states of $^{186}$Os were set in the Solotvina experiment with cadmium tungstate scintillator
enriched in $^{116}$Cd \cite{Dan03a}.

Here we present the final results of the experiment to search for double beta processes in zinc and tungsten
with the help of ZnWO$_4$ crystal scintillators. As a by-product of the experiment, we also have set
a new limit on $\alpha$ decay of $^{183}$W to the 375 keV metastable excited level of $^{179}$Hf.

\vspace{0.3cm}
\section{Experiment and data analysis}

The low background experiments to search for double beta processes
in zinc and tungsten isotopes have been performed by using zinc
tungstate crystal scintillators. The scintillation detectors with
ZnWO$_4$ crystals, the experimental set-up, the measurements and
the data analysis are described in detail in
\cite{Bel08,Bel09,Bel10}. Here we outline the main features of the
experiment.

\vspace{0.3cm}
\subsection{ZnWO$_4$ crystal scintillators}

Four ZnWO$_4$ crystal scintillators were used in our studies. Two crystals (117 g, $20\times19\times40$~mm,
and 699 g,  $\oslash44\times55$~mm) were produced by the Czochralski method \cite{Nag08,Nag09}
in the Institute for Scintillation Materials (Kharkiv, Ukraine). After 2130 h of low-background
measurements the crystal of 699 g was re-crystallized with the aim to study the effect of the re-crystallization
on the radioactive contamination of the material. The third ZnWO$_4$ crystal (141 g, $\oslash27\times33$~mm, the sample
had slightly irregular shape) was obtained by the re-crystallization process and used in further
measurements. The fourth ZnWO$_4$ crystal scintillator (239 g,  $\oslash41\times27$~mm) was produced
in the Nikolaev Institute of Inorganic Chemistry (Novosibirsk, Russia) by the
low-thermal gradient Czochralski technique \cite{Gal09a,Gal09b}.
The radioactive contaminations of the used crystals are reported in \cite{Bel10}.

\vspace{0.3cm}
\subsection{Low-background measurements}

The ZnWO$_4$ crystal scintillators were fixed inside a cavity of  $\oslash49\times59$~mm
in the central part of a cylindrical polystyrene light-guide of  $\oslash66\times 312$ mm.
The cavity was filled up with high purity silicone oil. The light-guide was optically connected
on opposite sides by optical couplant to two low radioactivity EMI9265-B53/FL 3'' photomultipliers (PMT).
The light-guide was wrapped by PTFE reflection tape. The detector was modified at the final stages of the
experiment: two polished quartz light-guides ($\oslash66\times100$~mm) were installed between the
polystyrene light-guide and the PMTs to suppress $\gamma$ ray background from the PMTs.

The detector has been installed in the low background DAMA/R\&D
set-up at the underground Gran Sasso National Laboratories of the
INFN (Italy) at the depth of $\approx$ 3600 m w.e. It was
surrounded by Cu bricks and sealed in a low radioactive air-tight
Cu box continuously flushed with high purity nitrogen gas (stored
deep underground for a long time) to avoid presence of residual
environmental radon. The outer passive shield consisted of 10 cm
of high purity Cu, 15 cm of low radioactive Boliden lead, 1.5 mm
of cadmium and 4/10 cm polyethylene/paraffin to reduce the
external background. The whole shield has been closed inside a
Plexiglas box, also continuously flushed by high purity nitrogen
gas. An event-by-event data acquisition system accumulates the
amplitude, the arrival time, and the pulse shape of the events.

The energy scale and the energy resolution of the ZnWO$_4$ detectors have been measured with $\gamma$ sources
$^{22}$Na, $^{60}$Co, $^{133}$Ba, $^{137}$Cs, $^{228}$Th, and $^{241}$Am. The energy resolution of the detectors
(full width at half of maximum) was in the range of ($8.8-14.6$)\% for 662 keV $\gamma$ line of $^{137}$Cs.

\vspace{0.3cm}
\subsection{Interpretation of the background}

As an example, the energy spectrum accumulated over 4305 h with the 239~g ZnWO$_4$ crystal scintillator
in the low background set-up is shown in Fig.~1. The energy spectrum
accumulated over 2798 h in the same conditions with low ($\sim 10$~keV) energy threshold is
presented in Inset.
A few visible peaks in the spectrum can be ascribed to $\gamma$ quanta of naturally occurring radionuclides
$^{40}$K, $^{214}$Bi ($^{238}$U chain) and $^{208}$Tl ($^{232}$Th) from the materials of the set-up.
The presence of the peak with energy $\approx 50$ keV can be explained by internal contamination of
the crystal by $^{210}$Pb. A comparatively wide peculiarity at the energy $\approx 0.8$ MeV is
mainly due to the $\alpha$ active nuclides of U and Th chains present in the crystal as trace contamination.

\begin{figure}[htb]
\begin{center}
\mbox{\epsfig{figure=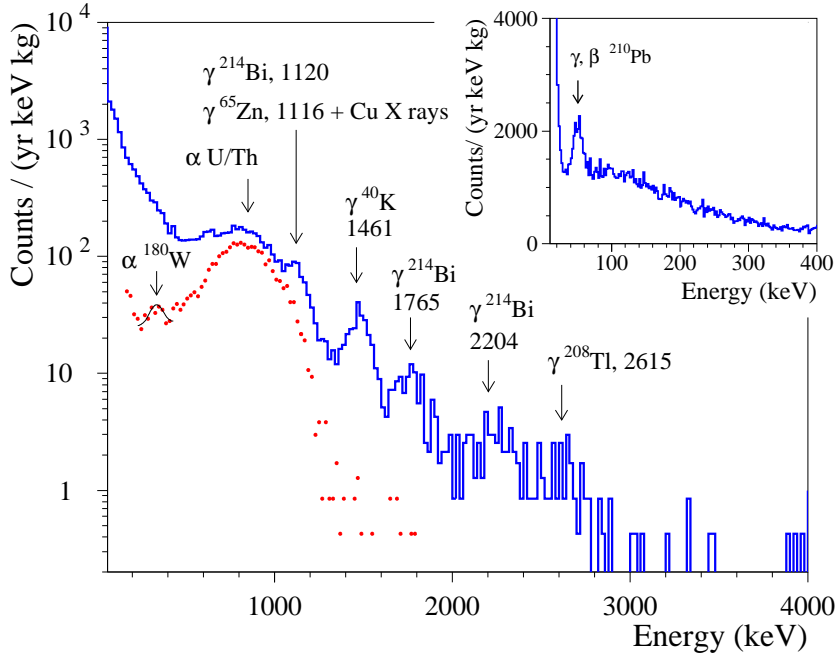,height=9.0cm}}
\caption{(Color online) The energy spectrum accumulated with the ZnWO$_4$ crystal
scintillator $\oslash41\times27$~mm in the low background DAMA/R\&D set-up over 4305 h.
The energy spectrum of $\alpha$ events selected by the pulse-shape discrimination is drawn
by points. Fit of the $\alpha$ peak of $^{180}$W by Gaussian function (solid line) is shown.
 (Inset) The energy spectrum of $\gamma$
and $\beta$ events selected by the pulse-shape discrimination technique from the data
measured over 2798 h with the same crystal scintillator in the set-up with lower energy
threshold and with additional quartz light-guides. Energies of $\gamma$ lines are in keV.}
\end{center}
\end{figure}

\vspace{0.2cm}
The radiopurity of the ZnWO$_4$ scintillators was already estimated \cite{Bel09,Bel10} by using the data of
the low-background measurements. The time-amplitude analysis (see details in \cite{Dan95,Dan01}),
the pulse-shape discrimination between $\beta$($\gamma$) and $\alpha$ particles \cite{Gat62},
the pulse-shape analysis of the double pulses (overlapped Bi-Po events) \cite{Dan03a,Dan03b,Bel07},
and the Monte Carlo simulation of the measured energy spectra were used to determine radioactive
contamination of the ZnWO$_4$ crystals. The radioactive contamination of the ZnWO$_4$ crystals is on
the level of $0.002-0.8$~mBq/kg (depending on the source); the total $\alpha$ activity is in the
range $0.2-2$~mBq/kg. Moreover, particular contaminations associated with the composition of
ZnWO$_4$ detector were observed \cite{Bel10}: the $EC$ active cosmogenic (or/and created by neutrons) nuclide
$^{65}$Zn ($T_{1/2} = 244.26$~d \cite{Fir98}) with activity $0.5-0.8$ mBq/kg
(depending on the ZnWO$_4$ sample) and the $\alpha$ active tungsten isotope $^{180}$W (with half-life:
$T_{1/2} \approx 10^{18}$~yr \cite{Bel10,Dan03b,Coz04,Zde05}, and energy of the decay:
$Q_{\alpha} = 2508(4)$~keV \cite{Aud03}) with activity 0.04 mBq/kg (see Fig.~1).

\vspace{0.2cm}
\section{Results and discussion}

\subsection{Response of the ZnWO$_4$ detectors to 2$\beta$ processes in zinc and tungsten}

The response functions of the ZnWO$_4$ detectors for the 2$\beta$  processes in Zn and W isotopes
were simulated with the help of the GEANT4 package \cite{GEANT} with the Low Energy Electromagnetic
extensions. The initial kinematics of the particles emitted in the decays was generated with
the DECAY0 event generator \cite{Decay0}. As examples, the expected energy distributions for
the ZnWO$_4$ detector $\oslash44\times55$~mm are shown in Fig.~2 and Fig.~3. The background models
included the internal contamination of the ZnWO$_4$ scintillators ($^{40}$K, $^{60}$Co, $^{65}$Zn, $^{87}$Rb,
$^{90}$Sr-$^{90}$Y, $^{137}$Cs, active nuclides from U/Th families), and the external $\gamma$ rays from
radioactive contamination of the PMTs and the copper box ($^{40}$K, $^{232}$Th, $^{238}$U); they were also
simulated with the help of the GEANT4 and DECAY0 packages.

\begin{figure}[htb]
\begin{center}
\mbox{\epsfig{figure=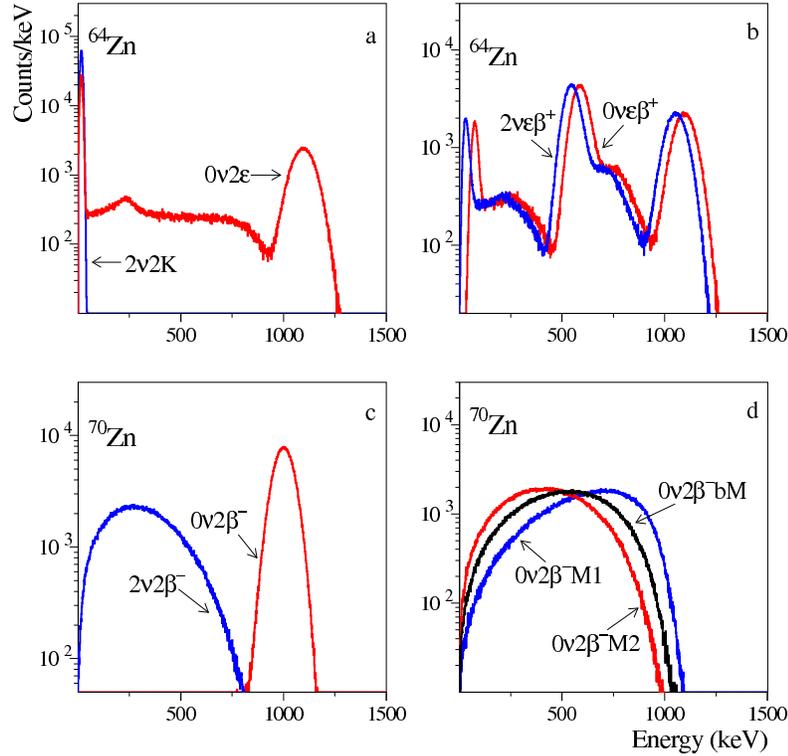,height=10.0cm}} \caption{(Color online)
Simulated response functions of the detector based on the ZnWO$_4$ scintillator  $\oslash44\times55$~mm
for the different 2$\beta$ processes in Zn isotopes: (a) 2$\varepsilon$ capture in $^{64}$Zn;
(b) $\varepsilon\beta^+$ decay of $^{64}$Zn; (c) 2$\beta^-$ decay of $^{70}$Zn; (d) $0\nu2\beta^-$  processes
with majorons emissions in $^{70}$Zn. One million decays was simulated for each process.}
\end{center}
\end{figure}

\begin{figure}[htb]
\begin{center}
\mbox{\epsfig{figure=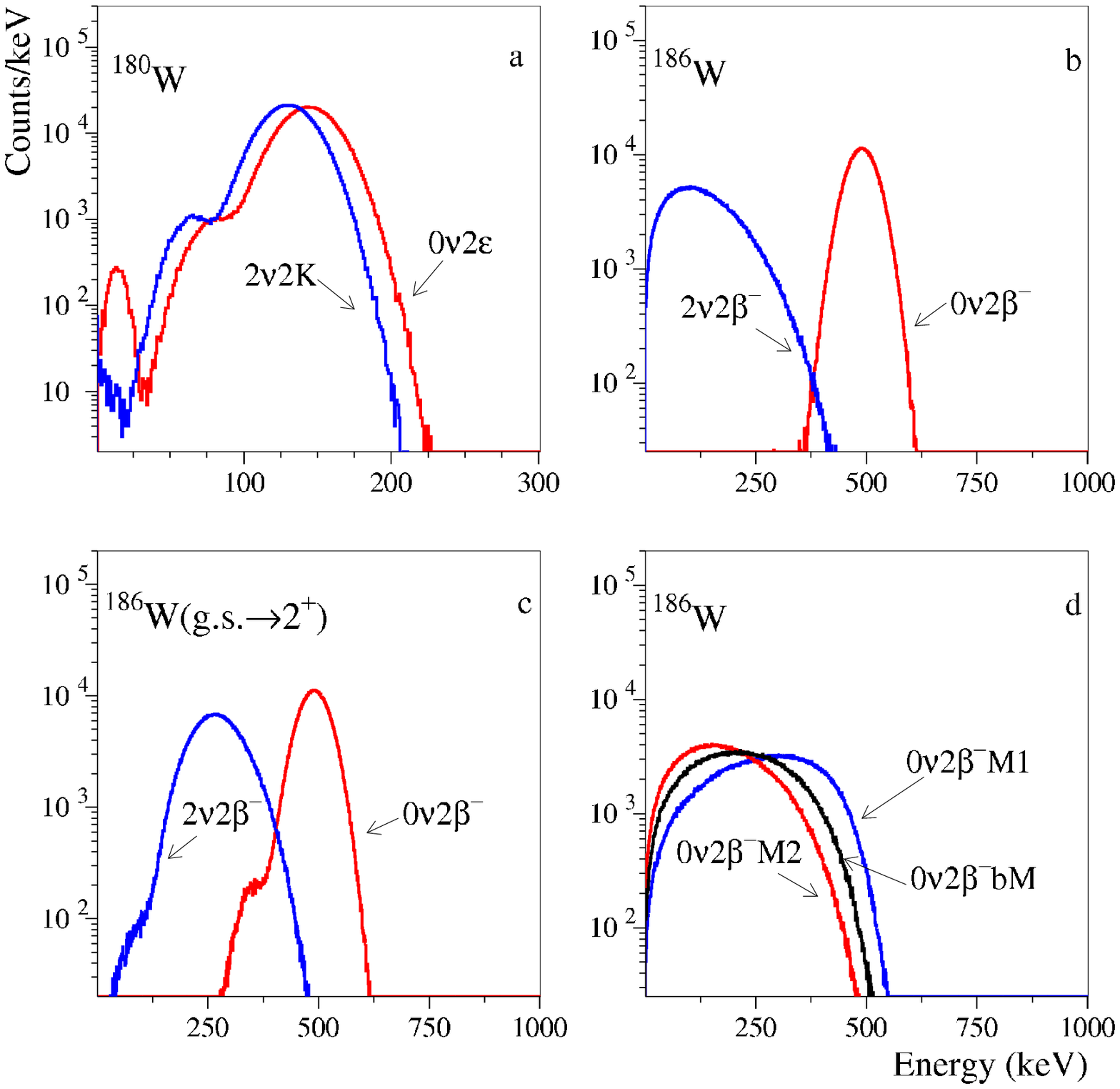,height=10.0cm}} \caption{(Color online)
Simulated response functions of the ZnWO$_4$ detector $\oslash44\times55$~mm for the different
2$\beta$ processes in W isotopes: (a) 2$\varepsilon$ capture in $^{180}$W;
(b) and (c) 2$\beta^-$ decay of $^{186}$W to the ground and excited states of $^{186}$Os, respectively;
(d) $0\nu2\beta^-$ decays of $^{186}$W with majorons emissions.
One million decays was simulated for each process.}
\end{center}
\end{figure}

\subsection{Double $\beta$ processes in $^{64,70}$Zn and $^{180,186}$W}

Comparing the simulated response functions with the measured energy spectra of the ZnWO$_4$ detectors, we have not found clear peculiarities, which can be evidently attributed to double beta decay of zinc or tungsten isotopes. Therefore only lower half-life limits can be set according to the formula:

\begin{equation}
\lim T_{1/2} = N \cdot \eta \cdot t \cdot \ln 2 / \lim S,
\end{equation}

\noindent where $N$ is the number of potentially 2$\beta$ unstable nuclei in a crystal scintillator,
$\eta$ is the detection efficiency, $t$ is the measuring time, and
$\lim S$ is the number of events of the effect searched for which can be excluded at a
given confidence level (C.L.; all the limits in the present study are given at 90\% C.L.).

For the 2$\nu$ double electron capture in $^{64}$Zn from the $K$ shell, the total energy released
in the detector is equal to 2$E_K = 16.7$ keV (where $E_K$ is the binding energy of
electrons on the $K$ shell of nickel atoms). The detection of such a small energy deposit requires
rather low energy threshold. In our measurements with the ZnWO$_4$ crystal scintillator $\oslash41\times27$~mm
the energy threshold of 10 keV was enough low (see Fig.~1, Inset) to observe at least
the higher energy part of the 2$\nu2K$ peak. Moreover, the background level (which is mainly due
to PMT noise in the low energy region) was decreased in comparison to our first measurement
\cite{Bel08} thanks to the improved scintillation properties of the ZnWO$_4$ crystal (slightly
higher transmittance, light output and energy resolution) and the enhanced light collection from the
scintillator. The light collection was increased by special treatment of the crystal surface,
which was diffused with the help of grinding paper (in our first experiment, the ZnWO$_4$ crystal scintillator
was polished \cite{Bel08}). Finally, a significant difference of ZnWO$_4$ pulse-shape (effective decay time is
$\approx 24$ $\mu s$ \cite{Dan05}) in comparison to much faster PMT noise (few nanoseconds)
offers the possibility to exploit the rejection
of residual PMT noise by using the pulse-shape discrimination.
However, this procedure eliminates some part of scintillation signals near energy threshold.
The energy dependence of the detection efficiency was determined with the help of $^{133}$Ba,
$^{137}$Cs, $^{228}$Th, and $^{241}$Am radioactive sources. The measured efficiency ranges from about
55\% at 15 keV up to about 95\% at 30 keV (one can compare these values with the detection
efficiencies 30\% at 15 keV and 65\% at 30 keV obtained in \cite{Bel08}).

\begin{figure}[htb]
\begin{center}
\mbox{\epsfig{figure=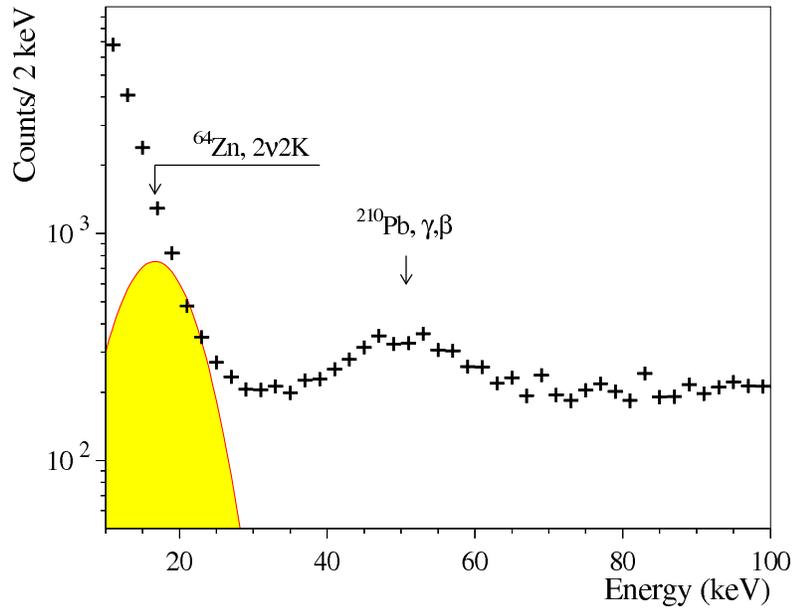,height=8.0cm}} \caption{
The energy spectrum of the ZnWO$_4$ crystal scintillator $\oslash41\times27$~mm measured over 2798 h,
corrected for the energy dependence of detection efficiency, together with the 2$\nu2K$ peak
of $^{64}$Zn with $T^{2\nu2K}_{1/2} = 1.1 \times 10^{19}$~yr excluded at 90\% C.L.}
\end{center}
\end{figure}

To set a limit on the 2$\nu2K$ decay of $^{64}$Zn, taking into account the proximity of the energy
threshold and the contribution from remaining PMT noise, we use a conservative requirement: the theoretical
energy distribution should not exceed the experimental one in any energy interval,
including error bars in the experimental values (see Fig.~4). In this way the limit
on the peak area is $\lim S$ = 4665 counts. Taking this value (already corrected
for the efficiency) for the peak area,
we conservatively give the following half-life limit on the 2$\nu2K$ process:

\begin{center}
$T^{2\nu2K}_{1/2}$($^{64}$Zn) $\geq 1.1 \times 10^{19}$~yr.
\end{center}

To estimate limits on other double $\beta$  processes, we have
used the following approach: the energy spectrum was fitted in the
energy range of an expected 2$\beta$ signal by a model built by
the simulated distributions of internal and external background
and of the effect searched for. The background model was composed
of $^{40}$K, $^{65}$Zn, $^{90}$Sr-$^{90}$Y, $^{137}$Cs, U/Th
inside a crystal (for fit of a low energy part of the data we have
also used a model of internal $^{87}$Rb), and $^{40}$K,
$^{232}$Th, $^{238}$U in the PMTs and the copper box. The
activities of the U/Th daughters in the crystals have been
restricted taking into account the data on the radioactive
contamination of the ZnWO$_4$ crystal scintillators \cite{Bel10}.
The initial values of the $^{40}$K, $^{232}$Th and $^{238}$U
activities inside the PMTs have been taken from Ref. \cite{Ber99},
while activities inside the copper box have been assumed to be
equal to the estimations obtained in Ref. \cite{Gun97}. We have
used different combinations of the accumulated data to reach the
maximal sensitivity to the double beta processes searched for.
Additionally we have also applied the so called 1$\sigma$ approach
when a statistical uncertainty of the number of events accumulated
in the energy region of the expected 2$\beta$ signal (square root
of the number of events) was taken as $\lim S$. This simple method
allows to obtain a correct evaluation of the experimental
sensitivity to the 2$\beta$  process searched for. It should be
stressed that the detection efficiencies in all the distributions
analyzed are at least 99.9\% for all the processes. Taking into
account the efficiency of  $\gamma$($\beta$) events selection by
the pulse-shape discrimination (98\%), the total detection
efficiencies are at least 97.9\% for all the 2$\beta$ processes
searched for.

Let us give an example of the analysis by using the two approaches to search for electron
capture with positron emission in $^{64}$Zn. 14922 events were observed in the energy interval
$530-1190$ keV of the spectrum accumulated with an exposure 0.3487 kg $\times$ yr (see Fig. 5),
which gives $\lim S$ = 122 counts. With the detection efficiency in the energy interval
to the 2$\nu\varepsilon\beta^+$  decay of $^{64}$Zn (82\%), one obtains the half-life limit
$T^{2\nu\varepsilon\beta^+}_{1/2} \geq 1.5 \times 10^{21}$~yr at 68\% C.L.
In order to apply the second approach, the starting and final energies of the fit were
varied as $380-550$ keV and $1260-1430$ keV, respectively, with the step of 10 keV.
The result of the fit in the energy region $520-1350$ keV was chosen as final giving the minimal
value of $\chi^2/n.d.f.$ = 119/98 = 1.23. It gives the total area of the 2$\nu\varepsilon\beta^+$
effect ($-208 \pm 254$) counts which corresponds (in accordance with the Feldman-Cousins procedure
\cite{Feld98}) to $\lim S$ = 238 counts in the full energy distributions for
2$\nu\varepsilon\beta^+$ decay. Thus, one can calculate the following
half-life limit, rather similar to the value obtained by using the 1$\sigma$ approach:

\begin{figure}[htb]
\begin{center}
\mbox{\epsfig{figure=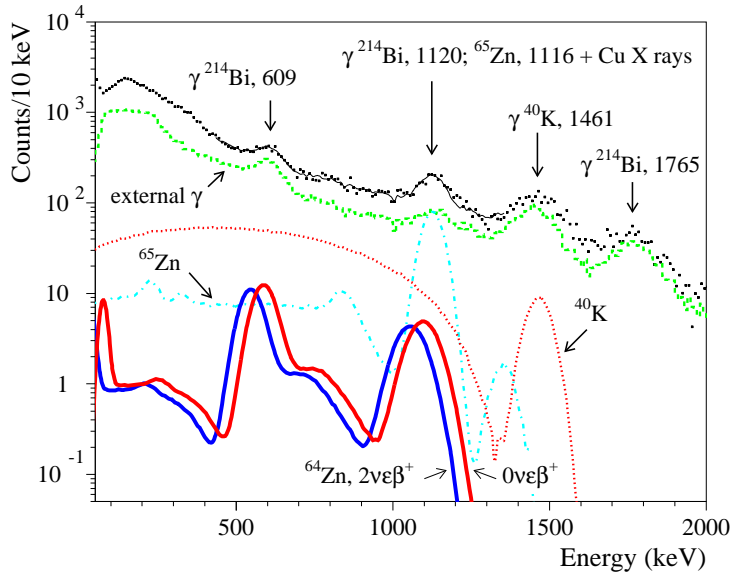,height=8.0cm}} \caption{(Color online)
The measured energy spectrum of the ZnWO$_4$ scintillation crystals (the total exposure
is 0.349 kg $\times$ yr) together with the GEANT4-simulated response functions for
$\varepsilon\beta^+$ process in $^{64}$Zn excluded at 90\% C.L.
The most important components of the background are shown.
The energies of $\gamma$ lines are in keV.}
\end{center}
\end{figure}

\begin{center}
$T^{2\nu\varepsilon\beta^+}_{1/2}$($^{64}$Zn) $\geq 9.4 \times 10^{20}$~yr.
\end{center}

In case of the neutrinoless electron capture with positron emission, the spectrum with
the total exposure 0.3487 kg $\times$ yr was fitted in the energy interval ($410-1370$) keV
($\chi^2/n.d.f.$ = 113/94 = 1.2). The fit gives the area of the effect searched for as
($52 \pm 129$) counts, which corresponds (again in accordance with the Feldman-Cousins procedure)
to $\lim S$ = 264 events. It allows to set the following limit on the half-life of
$^{64}$Zn relatively to the 0$\nu\varepsilon\beta^+$ decay:

\begin{center}
$T^{0\nu\varepsilon\beta^+}_{1/2}$($^{64}$Zn) $\geq 8.5 \times 10^{20}$~yr.
\end{center}
The energy distributions expected for the 2$\nu\varepsilon\beta^+$ and
0$\nu\varepsilon\beta^+$ decay of $^{64}$Zn, excluded at 90\% C.L., are shown in Fig.~5.

In case of $0\nu2\varepsilon$ decay of $^{64}$Zn, different particles are
emitted: X rays and Auger electrons from deexcitations in atomic shells,
$\gamma$ quanta and/or conversion electrons from
deexcitation of daughter nucleus. We suppose here that only one
$\gamma$ quantum is emitted in the nuclear deexcitation process;
it is the most pessimistic scenario from the point of view of
registration of such an event in a peak of full absorption at the
$Q_{\beta\beta}$ energy.
Unfortunately, 2$K$, $KL$, 2$L$ (and other) modes are not energetically resolved
in the high energy region due to finite energy resolution of the ZnWO$_4$ detectors.
So, the fit of the measured spectrum (exposure 0.3647 kg $\times$ yr) in the energy interval
$440-1350$ keV ($\chi^2/n.d.f.$ = 98/89 = 1.1) gives the area of the $0\nu2\varepsilon$ effect
searched for as ($-780 \pm 853$) counts. Taking into account the Feldman-Cousins procedure, we
calculated $\lim S$ = 742 events and the following limit on $0\nu2\varepsilon$
transition of $^{64}$Zn to ground state of $^{64}$Ni:

\begin{center}
$T_{1/2}^{0\nu2\varepsilon}$($^{64}$Zn) $\geq 3.2\times10^{20}$~yr.
\end{center}

Limits on double electron capture in $^{180}$W were set by analyzing all the data accumulated
in the experiment over 0.529 kg $\times$ yr. The low energy part of the spectrum is shown in Fig.~6.
The least squares fit of this spectrum in the
$100-260$ keV energy interval gives ($141 \pm 430$) counts for the 2$\nu2K$ peak searched for
($\chi^2/n.d.f.$ = 5.39/5 = 1.08), providing no evidence for the effect. These numbers
lead to an upper limit of 846 counts. Taking into account the detection efficiency for
this process close to 98\%, one can calculate the half-life limit:

\begin{center}
$T^{2\nu2K}_{1/2}$($^{180}$W) $\geq 1.0 \times 10^{18}$~yr.
\end{center}
The same approach gives the limit for the neutrinoless 2$\varepsilon$ process in $^{180}$W:

\begin{center}
$T^{0\nu2\varepsilon}_{1/2}$($^{180}$W) $\geq 1.3 \times 10^{18}$~yr.
\end{center}

The expected energy distributions for $0\nu2\varepsilon$ and $2\nu2K$ decay of $^{180}$W
corresponding to the best previous restrictions obtained in the Solotvina experiment
\cite{Dan03a} with the help of low background cadmium tungstate crystal scintillators
are presented in Fig. 6. Advancement of the sensitivity in the present study was reached
thanks to the lower background of ZnWO$_4$ detectors in comparison to CdWO$_4$ where the counting
rate in the energy interval up to 0.4 MeV was caused mainly by the $\beta$  decay of $^{113}$Cd.
The $0\nu2\varepsilon$ decay of $^{180}$W is of particular interest due to the possibility
of resonant process \cite{Suj02,Suj04,Kri11}.

\begin{figure}[htb]
\begin{center}
\mbox{\epsfig{figure=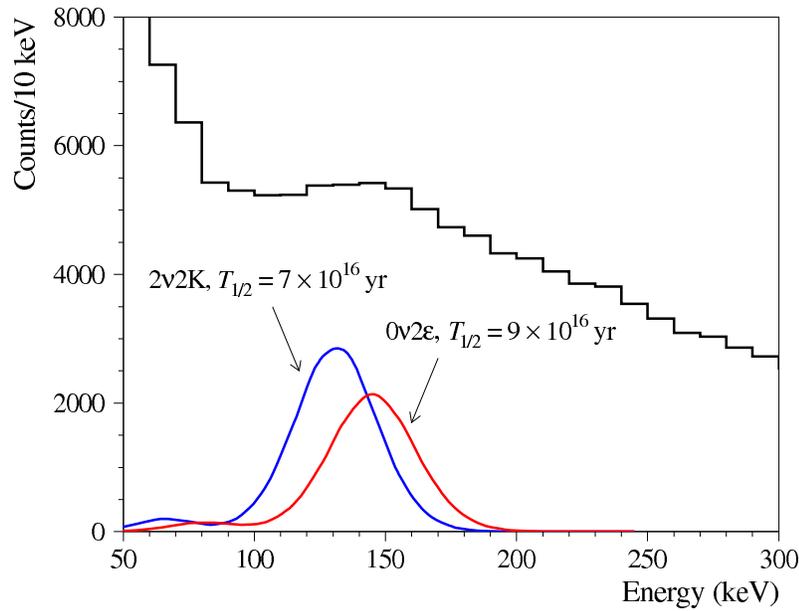,height=8.0cm}} \caption{(Color online)
Energy spectrum of the background of the ZnWO$_4$ detectors (exposure 0.529 kg $\times$ yr).
The simulated response functions for double electron capture in $^{180}$W are shown;
the half-lives $T^{2\nu2K}_{1/2} = 7 \times 10^{16}$ yr and
$T^{0\nu2\varepsilon}_{1/2} = 9 \times 10^{16}$ yr correspond to the best previous
limits obtained in \cite{Dan03a} with the help of cadmium tungstate crystal scintillators.}
\end{center}
\end{figure}

By using the approaches described above, the half-life limits on other 2$\beta$ decay
processes in $^{64}$Zn, $^{70}$Zn, and $^{186}$W have been obtained. All the results are
summarized in Table~2, where the data of the most sensitive previous
experimental investigations and theoretical estimations are given for comparison.

The obtained bounds are well below the existing theoretical predictions;
nevertheless most of the limits are higher than those established in previous experiments.
It should be stressed that in contrast to the results obtained in researches of double
$\beta^-$ decay (sensitivity of the best experiments is on the level of $10^{23}-10^{25}$~yr
\cite{2bTable,Bar10a,Bar10b}), only five nuclides ($^{40}$Ca \cite{Bel99}, $^{64}$Zn [present work],
$^{78}$Kr \cite{Gav06}, $^{112}$Sn \cite{Bar11}, and $^{120}$Te \cite{And11}) among 34 potentially
2$\varepsilon$, $\varepsilon\beta^+$, and 2$\beta^+$ active isotopes were investigated at the level
of sensitivity $\lim T_{1/2} \sim 10^{21}$~yr.

%%%\newpage
%%%\clearpage

%%%\begin{sidewaystable}[htb]
\begin{table}[ht]
\caption{Half-life limits on 2$\beta$ processes in Zn and W isotopes and
comparison with the theoretical predictions. Quoting best previous experimental results,
we exclude limits obtained on previous stages of our experiment \cite{Bel08,Bel09}.}
\begin{center}
\resizebox{0.96\textwidth}{!}{
\begin{tabular}{|l|l|l|l|l|l|}
\hline
 Transition & Decay & Level & \multicolumn{2}{c|}{ Experimental limits on $T_{1/2}$, yr at 90\% C.L.} &
 Theoretical estimations of \\
 \cline{4-5}
 ~ & channel & of daughter & Present work & The best previous results & the half-lives $T_{1/2}$, yr  \\
 ~ & ~ & nucleus & ~ & ~ & ($\left\langle m_{\nu}\right\rangle$ = 1 eV for $0\nu2\beta$ decay)  \\
 \hline
 $^{64}$Zn $\to$ $^{64}$Ni  & $2\nu2K$                 & g.s. & $\geq1.1\times10^{19}$    & $\geq6.0\times10^{16}$ \cite{Kie03}          & ($1.9-7.1$) $\times10^{26}$ \cite{Dom05} \\
 ~                          & ~                        & ~    & ~                         &  ~                                           & ($1.2\pm0.2)\times 10^{25}$ \cite{Gre08} \\
 ~                          & $0\nu2\varepsilon$       & g.s. & $\geq3.2\times10^{20}$    & $\geq7.4\times10^{18}$ \cite{Wil08}          & $-$  \\
 \cline{2-6}
 ~                          & 2$\nu\varepsilon\beta^+$ & g.s. & $\geq9.4\times10^{20}$    & = ($1.1\pm0.9$) $\times10^{19}$ \cite{Bik95} & ($0.9-2.2$) $\times10^{35}$ \cite{Dom05}  \\
 ~                          & ~                        & ~    & ~                         & $\geq1.3\times10^{20}$ \cite{Kim07}          & ($4.7\pm0.9)\times 10^{31}$ \cite{Gre08} \\
% ~ & ~ & ~ & ~ & $\geq1.3\times10^{20}$ \cite{Kim07} & ~  \\
 ~ & 0$\nu\varepsilon\beta^+$ & g.s. & $\geq8.5\times10^{20}$ & $\geq1.3\times10^{20}$ \cite{Kim07} & $-$  \\
  \hline
 $^{70}$Zn $\to$ $^{70}$Ge & $2\nu2\beta^-$ & g.s. & $\geq3.8\times10^{18}$ & $\geq1.3\times10^{16}$ \cite{Dan05} &
 $4.5\times10^{21}-3.6\times10^{24}$ \cite{Sta90}  \\
 ~ & ~ & ~ & ~ & ~ & $2.5\times10^{21}-6.4\times10^{23}$ \cite{Bob00}  \\
 ~ & ~ & ~ & ~ & ~ & $7.0\times10^{23}$ \cite{Dom05}  \\
 ~ & ~ & ~ & ~ & ~ & $\geq 3.1\times10^{22}$ \cite{Suh11}  \\
 ~ & $0\nu2\beta^-$ & g.s. & $\geq3.2\times10^{19}$ & $\geq7.0\times10^{17}$ \cite{Dan05} &
 $9.8\times10^{25}$ \cite{Sta90}  \\
 ~ & $0\nu2\beta^-M1$ & g.s. & $\geq6.0\times10^{18}$ & $-$ & $-$  \\
 ~ & $0\nu2\beta^-M2$ & g.s. & $\geq4.7\times10^{18}$ & $-$ & $-$  \\
 ~ & $0\nu2\beta^-bM$ & g.s. & $\geq5.4\times10^{18}$ & $-$ & $-$  \\
 \hline
 $^{180}$W $\to$ $^{180}$Hf & $2\nu2K$ & g.s. & $\geq1.0\times10^{18}$ & $\geq7.0\times10^{16}$ \cite{Dan03a} & $-$  \\
 ~ & $0\nu2\varepsilon$ & g.s. & $\geq1.3\times10^{18}$ & $\geq9.0\times10^{16}$ \cite{Dan03a} &
 $2.5\times10^{24}-2.5\times10^{26}$ \cite{Suj02}  \\
 ~ & ~ & ~ & ~ & ~ & $3.3\times10^{27}-5.0\times10^{30}$ \cite{Suj04}  \\
 ~ & ~ & ~ & ~ & ~ & $3.0\times10^{22}-4.0\times10^{27}$ \cite{Kri11}  \\
 \hline
 $^{186}$W $\to$ $^{186}$Os & $2\nu2\beta^-$ & g.s. & $\geq2.3\times10^{19}$ & $\geq3.7\times10^{18}$ \cite{Dan03a} &
 $7.1\times10^{23}-1.2\times10^{25}$ \cite{Sta90}  \\
 ~ & ~ & ~ & ~ & ~ & $\geq6.1\times10^{24}$ \cite{Cas94}  \\
 ~ & $2\nu2\beta^-$ & $2^{+}_{1}$(137 keV) & $\geq1.8\times10^{20}$ & $\geq1.0\times10^{19}$ \cite{Dan03a} & $-$  \\
 ~ & $0\nu2\beta^-$ & g.s. & $\geq1.0\times10^{21}$ & $\geq1.1\times10^{21}$ \cite{Dan03a} &
 $6.4\times10^{24}$ \cite{Sta90}  \\
 ~ & $0\nu2\beta^-$ & $2^{+}_{1}$(137 keV) & $\geq9.0\times10^{20}$ & $\geq1.1\times10^{21}$ \cite{Dan03a} & $-$  \\
 ~ & $0\nu2\beta^-M1$ & g.s. & $\geq5.8\times10^{19}$ & $\geq1.2\times10^{20}$ \cite{Dan03a} & $-$  \\
 ~ & $0\nu2\beta^-M2$ & g.s. & $\geq1.1\times10^{19}$ & $-$ & $-$  \\
 ~ & $0\nu2\beta^-bM$ & g.s. & $\geq1.1\times10^{19}$ & $-$ & $-$  \\
\hline
% \multicolumn{6}{l}{$^a$~The half-life calculated for 2$\varepsilon$ process (independent on shell of daughter nuclei).}

\end{tabular}}
\end{center}
\end{table}
%%%\end{sidewaystable}

%%%\clearpage
%%%\newpage

\subsection{Search for $\alpha$ decay of tungsten isotopes}

In addition to the previous observation of the $\alpha$ decay $^{180}$W $ \rightarrow$ $^{176}$Hf
(g.s. to g.s. transition) with CdWO$_4$ and CaWO$_4$ detectors \cite{Dan03b,Coz04,Zde05},
this rare process was observed also in our data with ZnWO$_4$ scintillators with
$T_{1/2} = 1.3^{+0.6}_{-0.5} \times 10^{18}$~yr \cite{Bel10} (one can
also see the $\alpha$ peak of $^{180}$W in the $\alpha$ spectrum presented in Fig. 1).

Here we report a new limit on the $\alpha$ decay of $^{183}$W ($Q_{\alpha} = 1680(2)$ keV
\cite{Aud03}; $\delta$  = 14.31(4)\% \cite{Ber11}) to the $1/2^-$ metastable level of $^{179}$Hf
(375 keV, $T_{1/2}$ = 18.67 s \cite{Fir98}). The
search for this process has been performed by using the data
of all the runs with the ZnWO$_4$ detectors with the total exposure 0.5295 kg $\times$ yr.
The signature of such a transition is delayed $\gamma$ quanta after the emission of the $\alpha$ particle.
The expected distribution of the time intervals between the $\alpha$ and the $\gamma$ events should
correspond to $T_{1/2}$ = 18.67 s. The time-amplitude technique \cite{Dan95,Dan01} and the pulse-shape
discrimination method \cite{Gat62,Dan05} have been applied to search for the $\alpha$ decay.
Taking into account the $\alpha$/$\beta$
ratio\footnote{It is defined as ratio of $\alpha$ peak position in the energy scale
measured with $\gamma$ sources to the real energy of $\alpha$ particles.} ($\alpha$/$\beta \approx 0.17$)
for the ZnWO$_4$ scintillator \cite{Dan05}, we expect to observe the $\alpha$ peak of the $^{183}$W decay to
the $^{179}$Hf metastable level at the energy 220 keV in $\gamma$ scale, with energy
resolution FWHM$_{\alpha}$ = 62 keV. All the $\alpha$ events selected within $150-270$ keV
have been used as triggers,
while a time interval $0.1-60$ s (88.9\% of $^{179}$Hf$^*$ decays) and a $325-425$ keV energy
window have been set for the second $\gamma$ events (energy resolution for gammas at the energy
375 keV: FWHM$_{\gamma}$ = 64 keV). Ninety five pairs were selected
from the data. The fit of the distribution of the selected ``$\alpha$ events'' by a simple
model built by a first degree polynomial function (to describe the background) plus a Gaussian
(the $\alpha$ peak searched for) gives the area of the effect searched for as ($10.5 \pm 17.6$) counts,
which corresponds to $\lim S$ = 39.4 events.
The excited 375 keV level of $^{179}$Hf deexcites with emission of two $\gamma$ quanta of 161 keV
and 214 keV \cite{Fir98}. The efficiency to detect a peak at the total energy release of 375 keV
in  ZnWO$_4$ detectors was calculated with the GEANT4 \cite{GEANT};
it was equal 0.71 to 0.86 in dependence on the volume of the ZnWO$_4$ crystal.
The half-life limit was calculated according to the formula analogous to (1):

\begin{center}
$\lim T_{1/2} = \ln2 \cdot \eta_{PSD} \cdot \sum \eta \cdot N
\cdot t / \lim S$,
\end{center}

\noindent where $\eta_{PSD}$ is the efficiency of the pulse-shape
discrimination (37.2\%), $N$ is the number of $^{183}$W nuclei,
$\eta$ is the registration efficiency of the total energy release
of 375 keV, and $t$ is the time of measurements with specific
ZnWO$_4$ detector. In result, we set the following limit on the
half-life of the $\alpha$ decay of $^{183}$W to the metastable 375
keV excited level of $^{179}$Hf:

\begin{center}
$T^{\alpha}_{1/2}$ ($^{183}$W $\rightarrow$ $^{179}$Hf$^*$, 375 keV) $\geq 6.7 \times 10^{20}$~yr.
\end{center}

\begin{figure}[htb]
\begin{center}
\mbox{\epsfig{figure=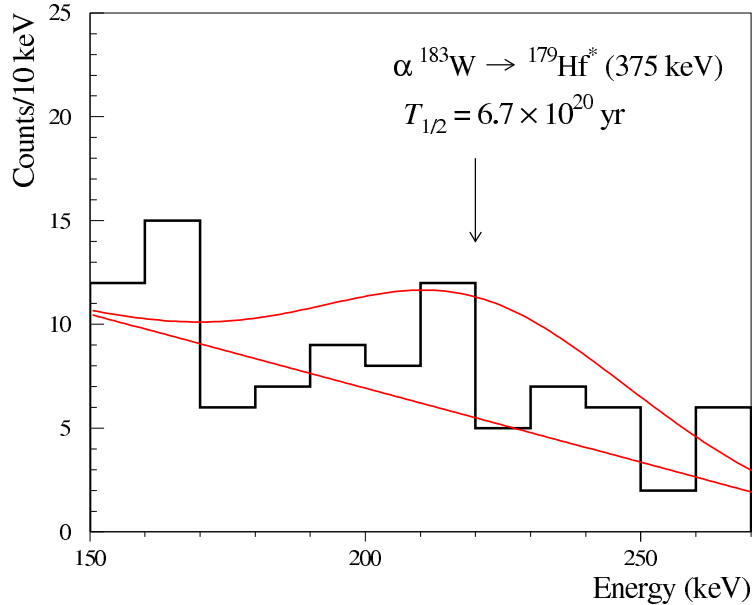,height=8.0cm}} \caption{(Color online)
Energy spectrum of the events selected by the time-amplitude and the pulse-shape analyses
from the data accumulated by ZnWO$_4$ detectors with an exposure 0.5295 kg $\times$ yr.
These events satisfy the search criteria for $\alpha$ transition of $^{183}$W to metastable
level of $^{179}$Hf. Polynomial function used as a background model and the
Gaussian peak corresponding to $\alpha$ decay of $^{183}$W with the half-life
$T_{1/2}$ = $6.7 \times 10^{20}$~yr excluded
at 90\% C.L. are also shown.}
\end{center}
\end{figure}

The energy spectrum of the selected events is shown in Fig. 7
together with the excluded $\alpha$ peak of $^{183}$W.

Despite the obtained limit is far away from the theoretical predictions
(f.i., $T_{1/2} \approx  1.3  \times 10^{50}$~yr \cite{Den09}), the limit is almost
two orders higher than the previous one $T_{1/2} \geq  1.0  \times 10^{19}$~yr
derived from the low background measurements
with a small (4.5 g) ZnWO$_4$ crystal scintillator \cite{Geo95}.

\section{Conclusions}

A low background experiment to search for 2$\beta$ processes in $^{64}$Zn, $^{70}$Zn,
$^{180}$W, and $^{186}$W was carried out over more than 19 thousands hours in the
underground Gran Sasso National Laboratories of the INFN by using radiopure
ZnWO$_4$ crystal scintillators. The total exposure of the experiment is 0.5295 kg $\times$ yr.

New improved half-life limits on double electron capture and electron capture
with positron emission in $^{64}$Zn have been set in the range: 10$^{19}$~yr to 10$^{21}$~yr
depending on the mode.
The indication on the ($2\nu + 0\nu$)$\varepsilon\beta^+$ decay of $^{64}$Zn with
$T_{1/2}$ = ($1.1 \pm 0.9$) $\times 10^{19}$~yr suggested in \cite{Bik95}
is completely disproved by the results of the present experiment.
Note that to date only four nuclides ($^{40}$Ca, $^{78}$Kr, $^{112}$Sn, and $^{120}$Te)
among 34 candidates to $2\varepsilon$, $\varepsilon\beta^+$, and 2$\beta^+$
processes were studied at similar level of sensitivity in direct experiments.
However, it is worth noting that the limits are still far from theoretical predictions.

In addition to $^{64}$Zn decays, in the course of the present experiment two important by-products were obtained:
(1) the new half-life limits on the 2$\beta$ processes in $^{70}$Zn, $^{180}$W, and $^{186}$W on
the level of $10^{18}-10^{21}$~yr
(the $0\nu2\varepsilon$ capture in $^{180}$W is of particular interest due to
the possibility of resonant process);
(2) rare $\alpha$ decay of $^{180}$W with a half-life
$T_{1/2} = 1.3^{+0.6}_{-0.5} \times 10^{18}$~yr has been observed and
new half-life limit on $\alpha$ transition of $^{183}$W to the 1/2$^-$ 375 keV metastable
level of $^{179}$Hf has been set as $T_{1/2} \geq 6.7 \times 10^{20}$~yr.

\section{Acknowledgments}

The group from the Institute for Nuclear Research (Kyiv, Ukraine)
was supported in part by the Project "Kosmomikrofizyka-2"
(Astroparticle Physics) of the National Academy of Sciences of
Ukraine. D.V. Poda was also partly supported by the Project
"Double beta decay and neutrino properties" for young scientists
of the National Academy of Sciences of Ukraine (Reg. No. 0110U004150).


\section*{References}
\begin{thebibliography}{99}

\bibitem{2bRev1} W.~Rodejohann, Neutrino-less double beta decay and particle physics, arXiv:1106.1334 [hep-ph],
                               submitted to Int. J. Mod. Phys. E;  \\
                                 F.T.~Avignone III, S.R.~Elliott, and J.~Engel, Rev. Mod. Phys. 80 (2008) 481; \\
                                 H.V.~Klapdor-Kleingrothaus, Int. J. Mod. Phys. E 17 (2008) 505.
\bibitem{2bRev2} H.~Ejiri, J. Phys. Soc. Japan 74 (2005) 2101;  \\
                                 F.T.~Avignone III, G.S.~King, and Yu.G.~Zdesenko, New J. Phys. 7 (2005) 6; \\
                                 S.R.~Elliot and J.~Engel, J. Phys. G: Nucl. Part. Phys. 30 (2004) R183; \\
                                 J.D.~Vergados, Phys. Rep. 361 (2002) 1; \\
                                 S.R.~Elliot and P.~Vogel, Ann. Rev. Nucl. Part. Sci. 52 (2002) 115; \\
                                 Yu.G.~Zdesenko, Rev. Mod. Phys. 74 (2002) 663.
\bibitem{Oscil} U.~Dore and D.~Orestano, Rep. Prog. Phys. 71 (2008) 106201; \\
                                R.N.~Mohapatra et al., Rep. Prog. Phys. 70 (2007) 1757.
\bibitem{Moh91} R.N.~Mohapatra and P.B.~Pal, \textit{Massive Neutrinos in Physics and Astrophysics}, 3rd ed., World
                              Sci., 2004.
\bibitem{Ber92} Z.G.~Berezhiani, A.Yu.~Smirnov, and J.W.F.~Valle, Phys. Lett. B 219 (1992) 99.
\bibitem{Moh88} R.N.~Mohapatra and E.~Takasugi, Phys. Lett. B 211 (1988) 192.
\bibitem{Bam95} P.~Bamert, C.P.~Burgess, and R.N.~Mohapatra, Nucl. Phys. B 449 (1995) 25.
\bibitem{Moh00} R.N.~Mohapatra, A.~Perenz-Lorenzana, and C.A.~de~Pires, Phys. Lett. B 491 (2000) 143.
\bibitem{2bTable} V.I.~Tretyak and Yu.G.~Zdesenko, At. Data Nucl. Data Tables 61 (1995) 43; 80 (2002) 83.
\bibitem{Bar10a} A.S.~Barabash, Phys. Rev. C 81 (2010) 035501.
\bibitem{Bar10b} A.S.~Barabash, Phys. At. Nucl. 73 (2010) 162.
\bibitem{Ack11} N.~Ackerman et al., arXiv:1108.4193v1 [nucl-ex].
\bibitem{Kla06} H.V.~Klapdor-Kleingrothaus and I.V.~Krivosheina, Mod. Phys. Lett. A 21 (2006) 1547.
\bibitem{GERDA} I.~Abt et al., hep-ex/0404039; \\
                                A.A.~Smolnikov and GERDA Collab., nucl-ex/0812.4194.
\bibitem{Majorana} Majorana Collab., nucl-ex/0311013; \\
                                C.E.~Aalseth et al., Nucl. Phys. B (Proc. Suppl.) 138 (2005) 217.
\bibitem{Mes01} A.P.~Meshik, C.M.~Hohenberg, O.V.~Pravdivtseva, and Ya.S. Kapusta, Phys. Rev. C 64 (2001) 035205.
\bibitem{Puj09} M.~Pujol, B.~Marty, P.~Burnard, and P.~Philippot, Geochim. Cosmochim. Acta 73 (2009) 6834.
\bibitem{Hir94} M.~Hirsch et al., Z. Phys. A 347 (1994) 151.
\bibitem{Cas94} O.~Castanos, J.G.~Hirsch, O.~Civitarese, and P.O.~Hess, Nucl. Phys. A 571 (1994) 276.
\bibitem{Win55} R.G.~Winter, Phys. Rev. 100 (1955) 142.
\bibitem{Vol82} M.B.~Voloshin, G.V.~Mitsel'makher, and R.A.~Eramzhyan, JETP Lett. 35 (1982) 656.
\bibitem{Ber83} J.~Bernabeu et al., Nucl. Phys. B 223 (1983) 15.
\bibitem{Suj02} Z.~Sujkowski and S.~Wycech, Acta Phys. Pol. B 33 (2002) 471.
\bibitem{Suj04} Z.~Sujkowski and S.~Wycech, Phys. Rev. C 70 (2004) 052501.
\bibitem{Kri11} M.I.~Krivoruchenko, F.~\v{S}imkovic, D.~Frekers, and A.~Faessler, Nucl. Phys. A 859 (2011) 140.
\bibitem{Aud03} G.~Audi, O.~Bersillon, J.~Blachot, and A.H.~Wapstra, Nucl. Phys. A 729 (2003) 337.
\bibitem{Ber11} M.~Berglund and M.E. Wieser, Pure Appl. Chem. 83 (2011) 397.
\bibitem{Bel08} P.~Belli et al., Phys. Lett. B 658 (2008) 193.
\bibitem{Bel09} P.~Belli et al., Nucl. Phys. A 826 (2009) 256.
\bibitem{Dan03a} F.A.~Danevich et al., Phys. Rev. C 68 (2003) 035501.
\bibitem{Bel10} P.~Belli et al., Nucl. Instr. Meth. A 626-627 (2010) 31.
\bibitem{Nag08} L.L.~Nagornaya et al., IEEE Trans. Nucl. Sci. 55 (2008) 1469.
\bibitem{Nag09} L.L.~Nagornaya et al., IEEE Trans. Nucl. Sci. 56 (2009) 2513.
\bibitem{Gal09a} E.N.~Galashov, V.A. Gusev, V.N. Shlegel, and Ya.V. Vasiliev, Func. Mat. 16 (2009) 63.
\bibitem{Gal09b} E.N.~Galashov et al., Crystallogr. Rep. 54 (2009) 689.
\bibitem{Dan95} F.A.~Danevich et al., Phys. Lett. B 344 (1995) 72.
\bibitem{Dan01} F.A.~Danevich et al., Nucl. Phys. A 694 (2001) 375.
\bibitem{Gat62} E.~Gatti and F.~De~Martini, Nuclear Electronics 2, IAEA, Vienna, 1962, p. 265.
\bibitem{Dan03b} F.A.~Danevich et al., Phys. Rev. C 67 (2003) 014310.
\bibitem{Bel07} P.~Belli et al., Nucl. Phys. A 789 (2007) 15.
\bibitem{Fir98} R.B.~Firestone et al., \textit{Table of Isotopes}, 8-th ed., John Wiley, New York,
                                1996 and CD update, 1998.
\bibitem{Coz04} C.~Cozzini et al., Phys. Rev. C 70 (2004) 064606.
\bibitem{Zde05} Yu.G.~Zdesenko et al., Nucl. Instr. Meth. A 538 (2005) 657.
\bibitem{GEANT} S.~Agostinelli et al., Nucl. Instr. Meth. A 506 (2003) 250; \\
                                J.~Allison et al., IEEE Trans. Nucl. Sci. 53 (2006) 270.
\bibitem{Decay0} O.A.~Ponkratenko et al., Phys. At. Nucl. 63 (2000) 1282;\\
                                    V.I.~Tretyak, to be published.
\bibitem{Dan05} F.A.~Danevich et al., Nucl. Instr. Meth. A 544 (2005) 553.
\bibitem{Ber99} R.~Bernabei et al., Il Nuovo Cim. A 112 (1999) 545.
\bibitem{Gun97} M.~G\"{u}nther et al., Phys. Rev. D 55 (1997) 54.
\bibitem{Feld98} G.J.~Feldman and R.D. Cousins, Phys. Rev. D 57 (1998) 3873.
\bibitem{Bel99} P.~Belli et al., Nucl. Phys. B 563 (1999) 97.
\bibitem{Gav06} Yu.M.~Gavrilyuk et al., Bull. Rus. Ac. Sci. Physics 75 (2011) 526.
\bibitem{Bar11} A.S.~Barabash et al., Phys. Rev. C 83 (2011) 045503.
\bibitem{And11} E.~Andreotti et al., Astropart. Phys. 34 (2011) 643.
\bibitem{Kie03} H.~Kiel, D.~Munstermann, and K.~Zuber, Nucl. Phys. A 723 (2003) 499.
\bibitem{Wil08} J.R.~Wilson et al., J. Phys. Conf. Ser. 120 (2008) 052048.
\bibitem{Dom05} P.~Domin, S.~Kovalenko, F.~\v{S}imkovic, and S.V.~Semenov, Nucl. Phys. A 753 (2005) 337.
\bibitem{Gre08} E.-W. Grewe et al., Phys. Rev. C 77 (2008) 064303.
\bibitem{Bik95} I.~Bikit et al., Appl. Radiat. Isot. 46 (1995) 455.
\bibitem{Kim07} H.J.~Kim et al., Nucl. Phys. A 793 (2007) 171.
\bibitem{Sta90} A.~Staudt, K.~Muto, and H.V.~Klapdor-Kleingrothaus, Europhys. Lett. 13 (1990) 31.
\bibitem{Bob00} A.~Bobyk, W.A.~Kaminski, and P.~Zareba, Nucl. Phys. A 669 (2000) 221.
\bibitem{Suh11} J.~Suhonen, Nucl. Phys. A 864 (2011) 63.
\bibitem{Den09} V.Yu.~Denisov and A.A.~Khudenko, Phys. Rev. C 79 (2009) 054614;  \\
                                Erratum Phys. Rev. C 82 (2010) 059901(E).
\bibitem{Geo95} A.Sh.~Georgadze et al., JETP Lett. 61 (1995) 882.

\end{thebibliography}
\end{document}